\documentclass[12pt]{article}

\date{\today}

\def\ren#1{\renewcommand{\arraystretch}{#1}}

\ren{1.5}

\begin{document}


\large \centerline{\bf Magnetic  QED
 }
\par
\vskip 0.5 truecm \normalsize
\begin{center}
{\bf C.~Ford}
\\
\vskip 0.5 truecm \it{
School of Mathematical Sciences\\
University College Dublin\\
 Belfield, Dublin 4\\Ireland\\
{\small\sf Christopher.Ford@ucd.ie}\\
}

\vskip 0.5 true cm

\end{center}

\vskip .7 truecm\normalsize
\begin{abstract}

A non-Hermitian form of QED is presented which describes interacting
Dirac monopoles.
   The theory is
related by a canonical
transformation to a model proposed by  Milton.
As in Hermitian QED  an abelian  gauge potential is coupled 
to a four-component
fermion. Under proper Lorentz transformations and time-reversal
the fermion field
transforms like a Dirac spinor but has a non-standard parity
transformation. This implements the property that magnetic charge,
unlike electric charge,  is parity-odd. A consequence of the
non-Hermiticity is that
 there is an
 attractive force between
identical charged particles, at least in the weakly coupled regime.
This effect can be understood even
at the classical level;
a simple calculation of the force between classical Dirac monopoles
is presented which shows that like charge monopoles attract and
opposite charges
repel.

\end{abstract}

\baselineskip=20pt
\bigskip
\bigskip
\medskip
\medskip
\medskip

This paper concerns the physical interpretation of non-Hermitian forms
of quantum electrodynamics (QED). In non-relativistic quantum mechanics
 some very simple non-Hermitian Hamiltonians
have been shown to exhibit a positive spectrum and unitary time-evolution
\cite{Bender:1998kf,Dorey:2001uw}.
These remarkable properties have also been identified in certain quantum
 field theories \cite{Kleefeld:2004jb,Kleefeld:2004qs,Bender:2005hf}.
Tentative steps have been taken to apply these ideas to gauge theory.
In particular,  Milton \cite{Milton:2003ax}
has proposed a non-Hermitian
version of QED. Unlike standard QED, parity ${\cal P}$ and
time-reversal ${\cal T}$ are not symmetries of the theory. However,
the combined symmetry ${\cal PT}$ is respected and on this basis the
theory is expected to exhibit a real energy spectrum and unitary
time-evolution.
The theory is also asymptotically free.

In this paper an alternative Lagrangian for non-Hermitian QED is given.
This theory is symmetric under ${\cal P}$ and ${\cal T}$.
Under parity and time-reversal the field strength transforms
like the Maxwell dual field strength.
This, together with the transformation properties
of the current, 
 suggests that the theory is a magnetic form of QED; 
 the elementary fermions carry magnetic
rather than electric charge.
As the  magnetic current is anti-Hermitian the force between like charge
particles is {\it attractive } rather than repulsive.
In fact, it is possible to understand this effect even
at the classical level.
A simple, albeit formal,
 argument is given that shows that the force between two
like charge Dirac monopoles is indeed attractive.  Opposite charges
repel.
A construction of  free-field representations for the gauge and matter fields
is outlined. This analysis shows that the new theory is related by a
canonical transformation to that of  Milton.

Massless QED is based on the Lagrangian
(the metric is $\hbox{diag}(1,-1,-1,-1)$)
\begin{equation}\label{qed}
{\cal L}=-{1\over 4}F_{\mu\nu}
F^{\mu\nu}+i\bar\psi\gamma^\mu\partial_{\mu}\psi +e\bar \psi
\gamma^\mu A_{\mu}\psi,
\end{equation}
where $F_{\mu \nu}=\partial_\mu A_\nu-\partial_{\nu}A_{\mu}$.
Here $A_{\mu}$ is a $U(1)$ gauge potential
 and $\psi$ is a Dirac spinor.
The corresponding quantum theory
has a Hermitian Hamiltonian and
is symmetric under parity ${\cal P}$ and time-reversal ${\cal T}$.
 Milton considered the Lagrangian \cite{Milton:2003ax} \footnote{
In \cite{Milton:2003ax} a real representation for 
Dirac spinors was adopted. In this paper, as also in 
 \cite{Bender:2005zz}, a conventional complex representation is assumed.} 
\begin{equation}\label{Milton}
{\cal L}=-{1\over 4}G_{\mu\nu}
G^{\mu\nu}+i\bar\psi\gamma^\mu\partial_{\mu}\psi +ig\bar \psi
\gamma^\mu B_{\mu}\psi,
\end{equation}
with $G_{\mu\nu}=\partial_{\mu}B_{\nu}-\partial_{\nu}B_{\nu}$,
$B_{\mu}$ being an abelian gauge potential, $\psi$ a Dirac spinor
and $g$  a real coupling constant.
 The theory couples a gauge
potential $B_{\mu}$ (assumed to be Hermitian) to the anti-Hermitian
current
\begin{equation}\label{Miltoncurrent}
j^\mu=ig\bar \psi \gamma^\mu \psi.\end{equation}
 A consequence of
the anti-hermiticity is that the current has a non-standard
transformation law under the (anti-unitary) operation of
time-reversal
\begin{equation}\label{TonMcurrent}
{\cal T}j^0({\bf r},t){\cal T}^{-1}=-j^0({\bf r},-t),~~~~~~~~
{\cal T}{\bf j}({\bf r},t){\cal T}^{-1}={\bf j}({\bf r},-t).
\end{equation}
Assuming the gauge potential transforms in the usual way
under time reversal, ie.
\begin{equation}\label{TonB}
{\cal T}B_0({\bf r},t){\cal T}^{-1}=B_0({\bf r},-t),~~~~~~~~~~
{\cal T}{\bf B}({\bf r},t){\cal T}^{-1}=-{\bf B}({\bf r},-t),
\end{equation}
the theory is not symmetric with respect to time-reversal. The
author of \cite{Milton:2003ax}
 was, however,  seeking a ${\cal
PT}$-symmetric theory. To achieve this  the gauge potential
$B_{\mu}$ is required to have the pseudovector parity transformation
\begin{equation}\label{PonB}
{\cal P}B_0({\bf r},t){\cal P}^{-1}=-B_0(-{\bf r},t),~~~~~~~~~~
{\cal P}{\bf B}({\bf r},t){\cal P}^{-1}={\bf B}(-{\bf r},t).
\end{equation}
The resulting theory is not parity-symmetric but
${\cal PT}$ is a symmetry of the theory.

Now consider the Lagrangian
\begin{equation}\label{newL}
{\cal L}=-{1\over 4}H_{\mu\nu} H^{\mu\nu}
+i\bar \lambda \gamma^\mu\partial_{\mu}
\lambda +ig \bar\lambda \gamma^\mu V_\mu
 \lambda.\end{equation}
Here $H_{\mu\nu}=\partial_{\mu}V_{\nu}-\partial_{\nu}V_{\mu}$ where
the gauge potential $V_{\mu}$
has  unconventional transformations
under both ${\cal P}$ and ${\cal T}$, that is
\begin{equation}\label{TonV}
{\cal T}V_0({\bf r},t){\cal T}^{-1}=-V_0({\bf r},-t),
~~~~~~~~~
{\cal T}{\bf V}({\bf r},t){\cal T}^{-1}={\bf V}({\bf r},-t), 
\end{equation}
and
\begin{equation}\label{PonV}
{\cal P}V_0({\bf r},t){\cal P}^{-1}
=-V_0(-{\bf r},t),~~~~~~~~
{\cal P}{\bf V}({\bf r},t){\cal P}^{-1}={\bf V}(-{\bf r},t).
\end{equation}
The spinor field $\lambda$
transforms like a Dirac spinor under proper Lorentz transformations
and time-reversal.
Under parity it transforms as
\begin{equation}\label{Ponlambda}
{\cal P}\lambda_\alpha({\bf r},t){\cal P}^{-1}
 ={P}_{\alpha \beta} \lambda_{\beta}^\dagger(-{\bf r},t),
\end{equation}
where $P_{\alpha \beta}$ denotes the matrix elements of the  Dirac matrix
$i\gamma^0 \gamma^2$
(here it is assumed that $\gamma_0=\gamma_0^T$ and $\gamma_2=\gamma_2^T$). This does not look
like a parity transformation; it is actually the standard form of
the ${\cal CP}$ transformation for Dirac spinor fields.
In much the same way that ${\cal CP}$
is unitary in standard QED (even though in
one-particle Dirac theory it is anti-unitary)
the above ${\cal P}$ transformation is unitary.
The theory couples the Hermitian gauge potential
$V_\mu$
to the anti-Hermitian current
\begin{equation}
\label{newcurrent}
j^\mu=ig\bar\lambda \gamma^\mu\lambda.\end{equation}
Under ${\cal P}$ and ${\cal T}$\footnote{Another
 anti-Hermitian current \cite{Bender:1999et}
with exactly
these transformation properties is $j^\mu=ig \bar \psi\gamma^\mu\gamma^5\psi$.
However, as charge conservation is spoiled by the chiral anomaly
it is not clear whether it leads to  a consistent quantum field theory.}
\begin{equation}
{\cal T}^{-1}j^0({\bf r},t){\cal T}=-j^0({\bf r},-t),~~~~~~~~
{\cal P}^{-1}{\bf j}({\bf r},t){\cal P}={\bf j}(-{\bf r},t).
\end{equation}
This non-Hermitian theory is symmetric under ${\cal T}$ and ${\cal P}$;
the non-standard transformation properties of $V_\mu$ compensate for those
of $j^\mu$. The field strength
 $H_{\mu\nu}$ transforms like the Maxwell dual field strength
and satisfies the ${\cal P}$ and ${\cal T}$ symmetric equation of motion
\begin{equation}\partial_{\mu} H^{\mu\nu}=j^\nu.\end{equation}
This indicates that the theory is a magnetic version of QED.
That magnetic charge is  parity odd just as electric
charge is ${\cal CP}$-odd `explains' the ${\cal CP}$-like
form of the parity transformation
for $V$ and $\lambda$.

It  remains to interpret the non-Hermiticity of the theory.
A consequence of the anti-hermiticity of the current is an
attractive force between identical charged particles, at least in
the weak coupling regime. In fact, this effect can  be understood
for \it classical \rm Dirac monopoles. In the absence of a Lorentz
force formula for magnetic charges it is  proposed to  compute the force
between two static monopoles using the electromagnetic energy
density formula
\begin{equation}
\label{EMenergy}
u={1\over 2}(E^2+B^2).\end{equation}
As the electric and magnetic fields enter this formula symmetrically
one might expect that the force between Dirac monopoles follows
exactly the same pattern as for electric charges. 
It is argued below that
 the Dirac string of the monopole breaks this symmetry leading
to an attractive force for like charges.

 A stationary Dirac monopole \cite{Dirac:1931}
centered at the origin
 can be described by the vector
potential
\begin{equation}\label{DiracA}
{\bf A}={g\over{4\pi r}}{y{\bf i}-x{\bf j}\over{(r-z)}},\end{equation}
where $g$ is the magnetic charge.
 In this gauge the Dirac string
lies on the positive $z$-axis. The magnetic field is
\begin{equation}
\label{DiracB}
{\bf B}=\nabla\times{\bf A}={g{\bf r}\over{4\pi r^3}}
-g\theta(z)\delta(y)\delta(z){\bf k}.
\end{equation}
 Consider two
Dirac monopoles with magnetic charge $g$
centred at the points
${\bf r}=a{\bf k}$ and ${\bf r}=-a{\bf k}$, respectively.
The total magnetic energy can be expressed formally as the integral
\begin{equation}
\label{FormalEnergy}
E={1\over 2}\int d^3x
\left({\bf B}_1\cdot{\bf B}_1+{\bf B}_2\cdot {\bf B}_2+2{\bf B}_1
\cdot{\bf B}_2\right),
\end{equation}
where ${\bf B}_1$ and ${\bf B}_2$ are the contributions to the magnetic
 field due to the first and second monopole, respectively.
Now
\begin{equation}
\label{InteractionEnergy}
U=
\int d^3x~ {\bf B}_1\cdot {\bf B}_2\end{equation}
is the part of the energy needed for the force computation since
the remainder
 comprises the infinite self energies
of the two monopoles which do not depend on the monopole separation.
Inserting the two magnetic fields into into (\ref{InteractionEnergy})
 gives
\begin{eqnarray}\label{Integral}
U&=&
{g^2\over{16\pi^2}}\int~d^3x \left(
{{\bf r}-a{\bf k}\over{|{\bf r}-a{\bf k}|^3}}\cdot
{{\bf r}+a{\bf k}\over{|{\bf r}+a{\bf k}|^3}}
 -4\pi\theta(z-a)\delta(x)\delta(y){\bf k}\cdot
{{\bf r}+a{\bf k}\over{|{\bf r}+a{\bf k}|^3}}\right. \\
&&~~~~~~~~~~~~~~~~~~~~~~~~~~~~~~\left. +4\pi
{{\bf r}-a{\bf k}\over{|{\bf r}-a{\bf k}|^3}}\cdot
\theta(-z-a)\delta(x)\delta(y){\bf k}\right).
\nonumber
\end{eqnarray}
Here ${\bf B}_1$ is a translation of (\ref{DiracB}) 
and 
\begin{equation}
{\bf B}_2={g\over{4\pi}}{{\bf r} +a{\bf k}\over{|{\bf r}+a{\bf k}|^3}}
+g\theta(-z-a)\delta(y)\delta(z){\bf k},
\end{equation}
so that the Dirac string of the second monopole lies on the
part of the negative axis below $z=-a$.
The first term in (\ref{Integral}) 
gives the expected
Coulomb repulsion, $g^2/(8\pi|a|)$.
Performing the other two integrals gives a contribution
twice the Coulomb term but with the opposite sign.
Accordingly,
\begin{equation}U=-{g^2\over{8\pi |a|}},\end{equation}
giving an attractive force between like charge monopoles.
Similarly, the force between opposite charges is repulsive.
In this computation we have taken the Dirac strings to lie on the $z$-axis.
However, the result is independent of the string placement
(provided the two strings do not intersect which would lead to
an ill-defined cross term in the $U$ integral).
A relativistic force law that incorporates the magnetic attraction 
property
is
\begin{equation}
m{d^2x_\mu\over{d\tau^2}}=\left(eF_{\mu\nu}-g\tilde F_{\mu\nu}\right)
{dx^\nu\over{d\tau}},
\end{equation}
where 
$\tau$ denotes the proper time and 
$\tilde F_{\mu\nu}$ is the
 Hodge dual of $F_{\mu\nu}$. Maxwell's equations are 
\begin{equation}
\partial_{\mu}F^{\mu\nu}=j_{e}^\nu,~~~~~~~~
\partial_{\mu}\tilde F^{\mu\nu}=j^\nu_{m},
\end{equation}
where $j_e^\mu$ and $j_m^\mu$ are the 
electric and magnetic currents,
respectively.

To conclude, the derivation of free-field
representations for the QED theories  is outlined.
It is instructive to start with the photon
field for ordinary QED (see for example \cite{Bjorken:1965}).
A free photon field (the gauge is fixed so that
 $A_0=0$ and $\nabla\cdot{\bf A}$)
takes the form
\begin{equation}
{\bf A}({\bf r},t)=\int
{d^3k\over{\sqrt{2\omega(2\pi)^3}}}\sum_{\lambda=1}^2
{\bf e}(k,\lambda)
\left[
a(k,\lambda)e^{-ik\cdot x}+
a^\dagger(k,\lambda) e^{ik\cdot x}\right],
\end{equation}
Here $k_0=\omega=|{\bf k}|$
so that $k_\mu k^\mu=0$.
The polarization vectors, ${\bf e}(k,\lambda)$ ~~$\lambda=1,2$,
 are orthogonal to
${\bf k}$,
and satisfy
\begin{equation}
{\bf e}(k,\lambda)\cdot {\bf e}(k',\lambda)=\delta_{\lambda \lambda'},
~~~~~~~~
{\bf e}(-k,1)=-{\bf e}(k,1),~~~~~~~~~
{\bf e}(-k,2)=+{\bf e}(k,2).
\end{equation}
The creation and annihilation operators obey
the commutation relations
\begin{equation}
[a(k,\lambda),a^\dagger(k',\lambda')]=\delta^3({\bf k}-{\bf k}')
\delta_{\lambda \lambda'},\
\end{equation}
and
\begin{equation}
[a(k,\lambda),a(k',\lambda')]=[a^\dagger(k,\lambda),
a^\dagger(k',\lambda')]=0.\end{equation}
The action of the discrete symmetries is as follows
\begin{equation}
{\cal T}^{-1}{\bf A}({\bf r},t){\cal T}=
-{\bf A}({\bf r},-t),~~~~~
{\cal P}^{-1}{\bf A}({\bf r},t){\cal P}=-{\bf A}(-{\bf r},t),
~~~~~
 ({\cal CP})^{-1}{\bf A}({\bf r},t){\cal CP}={\bf A}(-{\bf r},t).
\end{equation}
${\cal P}$ and ${\cal CP}$ have the representations
\begin{equation}
{\cal P}=\exp\left[-{i\pi\over2}\int
 d^3k\sum_{\lambda=1}^2\left(a^\dagger(k,\lambda)
a(k,\lambda)+
a^\dagger(k,\lambda)a(-k,\lambda)\right)\right],
\end{equation}
and
\begin{equation}
{\cal CP}=\exp\left[{i\pi\over2}\int
 d^3k\sum_{\lambda=1}^2\left(a^\dagger(k,\lambda)
a(k,\lambda)-
a^\dagger(k,\lambda)a(-k,\lambda)\right)\right].
\end{equation}

We require a  Hermitian quantum field $V_\mu$ satisfying the same
commutation relations as $A_\mu$
but with the \it opposite
\rm
transformations to $A_\mu$ under ${\cal T}$, ${\cal P}$ and
${\cal CP}$.
Consider
 \begin{equation}
{\bf V}({\bf r},t)=\int
{d^3k\over{\sqrt{2\omega(2\pi)^3}}}\sum_{\lambda=1}^2
{\bf e}(k,\lambda)
\left[
ia(k,\lambda)e^{-ik\cdot x}-i
a^\dagger(k,\lambda) e^{ik\cdot x}\right].
\end{equation}
This is the same as ${\bf A}({\bf r},t)$ but with
$a(k,\lambda)$ replaced by $ia(k,\lambda)$ and
$a^\dagger(k,\lambda)$ replaced by
$-i a^\dagger(k,\lambda)$,
a canonical transformation.
The $i$ insertions switch the time-reversal
 properties of the field so that (\ref{TonV}) holds.
Note that ${\cal P}$ and ${\cal CP}$ are unaffected so
 that (\ref{PonV}) is not satisfied.
However, all one needs to do is to swap ${\cal P}$ and ${\cal CP}$.
That is the parity operator ${\cal P}$ is \it defined
\rm
to be the standard form of ${\cal CP}$ and
${\cal CP}$ is defined to be the standard form of ${\cal P}$.
The same procedure yields a free-field representation
of the fermion $\lambda$;
simply take a standard Dirac fermion and exchange the definitions
of ${\cal P}$ and ${\cal CP}$. This provides a fermion field $\lambda$
with a standard ${\cal T}$ transform and the non-standard
parity transformation (\ref{Ponlambda}).
Interacting fields may be formally
defined in the usual way through
$V_\mu^{int}({\bf r},t)=e^{iHt}V_\mu({\bf r},0)e^{-iHt}$
and $\lambda^{int}({\bf r},t)
=e^{iHt}\lambda({\bf r},0)e^{-iHt}$.

To define $B_\mu$ in the Milton theory  take $B_\mu$ to be
$A_\mu$ and swap the definition of ${\cal P}$ and ${\cal CP}$.
For the fermion field the standard definitions of ${\cal P}$
and ${\cal CP}$ are retained. This is consistent since
for free fields the parity operator decomposes into
commuting gauge and fermionic pieces, ${\cal P}=
{\cal P}_{gauge}{\cal P}_{fermion}$; one can choose a non-standard `magnetic'
${\cal P}_{gauge}$ together with a standard `electric' ${\cal P}_{fermion}$.
In fact, taking the `non-standard' form
for both ${\cal P}_{gauge}$ and ${\cal P}_{fermion}$
gives a parity-symmetric theory. 
Then the fermion would transform like  $\lambda$ and one can write the
 Lagrangian as
\begin{equation}
{\cal L}=-{1\over 4}G_{\mu\nu}G^{\mu\nu}
+i\bar \lambda\gamma^\mu\partial_{\mu}\lambda
+ig\bar \lambda \gamma^\mu B_{\mu}\lambda,
\end{equation}
which is (\ref{newL}) with $V_\mu$ replaced by $B_\mu$.
As $B_\mu$ and $V_\mu$ are related by a canonical
transformation so are the two non-Hermitian theories.


\begin{thebibliography}{99}




\bibitem{Bender:1998kf}
C.~M.~Bender and S.~Boettcher, Phys. Rev. Lett. {\bf 80},
5243 (1998)

[arXiv:physics/9712001].





\bibitem{Dorey:2001uw}
 P.~Dorey, C.~Dunning and R.~Tateo,  
J. Phys. A {\bf 34}, 5679 (2001)
[arXiv:hep-th/0103051].





\bibitem{Kleefeld:2004jb}
F.~Kleefeld,
`Non-Hermitian quantum theory and it holomorphic representation:
Introduction and some Applications',
[arXiv:hep-th/0408028].


\bibitem{Kleefeld:2004qs}
F. Kleefeld, `Non-Hermitian quantum theory and its holomorphic representation:
Introduction and  applications',
[arXiv:hep-th/0408097].



\bibitem{Bender:2005hf}
C.~M.~Bender, H.~F.~Jones and R.~J.~Rivers,
Phys. Lett. B {\bf 625}, 333 (2005)
[arXiv:hep-th/0508105].




\bibitem{Milton:2003ax}
K.~A.~Milton, 
Czech. J. Phys. {\bf 54}, 85 (2004)
[arXiv:hep-th/0308035].


\bibitem{Bender:2005zz}
C.~M.~Bender, I.~Cavero-Pelaez, K.~A.~Milton
and K.~V.~Shajesh,
 Phys. Lett. B {\bf 613}, 97 (2005)
[arXiv:hep-th/0501180].


\bibitem{Bender:1999et}
C.~M.~Bender and K.~A.~Milton,
J. Phys. A {\bf 32}, L87 (1999).


\bibitem{Dirac:1931}
P.~A.~M.~ Dirac, Proc. Royal Society London A133 (1931) 60.



\bibitem{Bjorken:1965}
J. D. Bjorken and S. Drell, Relativistic Quantum Fields,
McGraw-Hill 1965.

\end{thebibliography}
\end{document}